\documentclass[prd,aps,preprint,groupedaddress,superscriptaddress,nofootinbib]{revtex4}

\usepackage{comment}
\usepackage[utf8]{inputenc} 
\usepackage{graphicx,color,overpic,mathtools}
\usepackage{amsthm,amsmath,amssymb,hyperref,mathrsfs}
\usepackage{braket,bm,bbm,setspace}
\PassOptionsToPackage{normalem}{ulem}
\usepackage{ulem} 
\usepackage{physics}
\usepackage{float}
\usepackage[makeroom]{cancel}
\usepackage[english]{babel}
\usepackage{graphicx}
\graphicspath{ {images/} }
\addto\captionsspanish{}
\hypersetup{
    colorlinks=false,
    pdfborder={0 0 0},
}

\begin{document}
\vspace*{-2cm}
\hfill YITP-22-53
\vspace*{1.1cm}
\title{Regular black holes without mass inflation instability}
\vspace*{-1.5cm}
\author{Ra\'ul Carballo-Rubio}
\affiliation{CP3-Origins, University of Southern Denmark, Campusvej 55, DK-5230 Odense M, Denmark}
\affiliation{Florida Space Institute, University of Central Florida, 12354 Research Parkway, Partnership 1, 32826 Orlando, FL, USA}
\author{Francesco Di Filippo}
\affiliation{Center for Gravitational Physics, Yukawa Institute for Theoretical Physics, Kyoto University, Kyoto 606-8502, Japan}
\author{Stefano~Liberati}
\affiliation{SISSA - International School for Advanced Studies, Via Bonomea 265, 34136 Trieste, Italy}
\affiliation{IFPU, Trieste - Institute for Fundamental Physics of the Universe, Via Beirut 2, 34014 Trieste, Italy}
\affiliation{INFN Sezione di Trieste, Via Valerio 2, 34127 Trieste, Italy}
\author{Costantino Pacilio}
\affiliation{Dipartimento di Fisica, ``Sapienza" Universit\`a di Roma \& Sezione INFN Roma1, Piazzale Aldo Moro 5, 00185, Roma, Italy}
\author{Matt Visser}
\affiliation{
School of Mathematics and Statistics, Victoria University of Wellington, PO Box 600, Wellington 6140, New Zealand
}

\begin{abstract}
Generic models of regular black holes have separate outer and inner horizons, both with nonzero surface gravity. It has been shown that a nonzero inner horizon surface gravity results in exponential instability at the inner horizon controlled by this parameter. This phenomenon takes the name of ``mass inflation instability'', and its presence has put in question the physical viability of regular black holes as alternatives to their (singular) general relativity counterparts. 
In this paper, we show that it is possible to make the inner horizon surface gravity vanish, while maintaining the separation between horizons, and a non-zero outer horizon surface gravity. We construct specific geometries satisfying these requirements, and analyze their behavior under differ\-ent kinds of perturbations, showing that the exponential growth characteristic of mass inflation instability is not present for these geometries. These ``inner-extremal'' regular black holes are thereby better behaved than singular black holes and generic regular black holes, thus providing a well-motivated alternative of interest for fundamental and phenomenological studies.

\end{abstract}


\maketitle


\def\R{{\mathbb{R}}}
\def\O{{\mathcal{O}}}

\null 
\vspace{-80pt}
\section{Introduction}

Regular black holes are deformations of solutions of the vacuum Einstein field equations in which the inner singularity is excised and replaced by a non-singular core. This procedure is highly non-unique, and the literature contains numerous proposals for both spherically symmetric~\cite{Bardeen1968,Dymnikova1992,Bonanno:2000ep,Hayward:2005gi,Modesto:2008im,Platania:2019kyx} and rotating~\cite{Bambi:2013ufa,Toshmatov:2014nya,Azreg-Ainou:2014pra,Eichhorn:2021etc,Eichhorn:2021iwq} regular black holes. The main distinguishing feature of these proposals is the existence of an inner horizon. In fact, in spherical symmetry it has been shown that inner horizons must be present for the regularity conditions to hold~\cite{Dymnikova:2001fb,Carballo-Rubio:2019fnb,Carballo-Rubio:2019nel}.\footnote{The question of whether inner horizons are necessary for regularity in rotating black holes is technically open, though all known proposals display this feature.}

Generically, inner horizons have a non-zero surface gravity, which translates into a characteristic dynamical behavior of geodesics and metric perturbations in their vicinity. The main effect of this dynamical behavior is a phase of exponential growth of the gravitational energy in a neighborhood of the inner horizon, which is known as mass inflation~\cite{Poisson1989,Ori:1991zz,Hamilton2008}. Mass inflation is present when both ingoing and outgoing perturbations are considered. One can employ simplified models of matter perturbations around the inner horizon while still capturing the salient features of mass inflation. Existing analyses either model both kinds of perturbations as null shells~\cite{Brown:2011tv,Carballo-Rubio:2018pmi,Bertipagani:2020awe}, or use null shells to describe outgoing perturbations while using null dust to characterize ingoing perturbations~\cite{Brown:2011tv,Bonanno:2020fgp,Carballo-Rubio:2021bpr}, the latter dating back to Ori's model of mass inflation of Reissner--Nordstr\"{o}m black holes \cite{Ori:1991zz}.

In this paper, we consider regular black holes with vanishing surface gravity at the inner horizon, and show that they do not display an exponential growth of perturbations, in contrast to the case of a finite surface gravity. At the same time, we show that the coefficients of the metric can be chosen in such a way that the outer horizon is well separated from the inner horizon and features a finite nonzero surface gravity. In other words, regular black hole metrics without mass inflation can be constructed while maintaining the outer horizon to be non-extremal. We propose the name ``inner-extremal'' regular black holes for these geometries.
The absence of unstable modes makes these regular black holes appealing candidates to describe astrophysical black holes, candidates which are well behaved down to the inner core.

The paper is organized as follows. We start {by} describing the geometries of inner-extremal regular black holes in Sec.~\ref{sec:geom}. We then study the behaviour of the metric under the two models of perturbations mentioned above, the double-null shell model and Ori's model, in Secs.~\ref{sec:dnull} and~\ref{sec:ori} respectively. We finish the paper with a summary of our results and a discussion of their implications in Sec.~\ref{sec:conc}.

\section{Inner-extremal regular black holes 
\label{sec:geom}}

The aim of this section is to show the existence of regular black hole geometries for which the surface gravity at the inner horizon vanishes. We will work in spherical symmetry for simplicity, and we start by analyzing static configurations without the accretion of matter. Under these assumptions, the most general line element can be written in advanced null coordinates as
\begin{equation}\label{eq:genmet}
       \text{d}s^2=-e^{-2\phi(r)}F(r)\text{d}v^2+2e^{-\phi(r)}\text{d}v\text{d}r+r^2\text{d}\Omega^2,
\end{equation}
where $F(r)$ and $\phi(r)$ are two arbitrary functions, with the only restriction that $\phi(r)$ must be finite for the metric determinant to be well defined.

It is also useful to introduce the Misner--Sharp quasi-local mass $m(r)$ defined by the expression \cite{Misner1964,Hayward1994}
\begin{equation}\label{eq:msdef}
    F(r)=1-\frac{2m(r)}{r}.
\end{equation}
For simplicity, we will restrict our attention to geometries in which $F(r)$ is a rational function of the radial coordinate, namely
\begin{equation}
    F(r)=\frac{N_n(r)}{D_n(r)},
\end{equation}
where $N_n$ and $D_n$ are polynomials of the same degree $n$. This simplifying assumption has been considered before, for instance in~\cite{Frolov:2016pav}.

The conditions for regularity at $r=0$ have been studied previously, e.g.~\cite{Carballo-Rubio:2019fnb}. If the metric functions are finite everywhere, so that we can write
\begin{align}
    m(r)&=m_0+m_1r+m_2r^2+\mathcal{O}(r^3),\nonumber\\
    \phi(r)&=\phi_0+\phi_1r+\phi_2r^2+\mathcal{O}(r^3),
\end{align}
then demanding regularity is equivalent to
\begin{align}\label{eq:metreg}
m_0=m_1=m_2=\phi_1=0.
\end{align}
These conditions can be obtained in different but equivalent ways, for instance calculating the effective energy density and pressures at $r=0$, or calculating the orthonormal components of the Riemann tensor at $r=0$. Demanding that these quantities are finite leads to Eq.~\eqref{eq:metreg}. Note further that, when including time dependence, the regularity conditions are the natural generalization of Eq.~\eqref{eq:metreg} with $m_0$, $m_1$, $m_2$ and $\phi_0$ now functions of $v$.

Eq.~\eqref{eq:metreg} implies that $F(r)=1$ at $r=0$. It is then clear that regularity implies that the metric must have an even number of horizons counting multiplicities, where the locations of the horizons are defined as usual by the roots of $F(r)$ \cite{Carballo-Rubio:2018pmi}. A black hole with a single simple root of $F(r)$ such as the Schwarzschild black hole cannot be regular. At the very least two simple roots, or one double root, are needed for regularity. These are the standard type of regular black holes analyzed in the literature. 

However, here we are interested in a different realization. We will consider situations in which $F(r)$ has two roots $r_+\geq r_-$. In most situations these two roots will be different, but it is useful to keep the analysis general enough so that the coincidence limit can be taken. The quantity $r_+$ thus indicates the position of the outer horizon, and $r_-$ the position of the inner horizon. The inner surface gravity is then given by
\begin{equation}
   \kappa_-=e^{-\phi(r_-)}\left.\frac{dF(r)}{dr}\right|_{r=r_-}.
\end{equation}
For situations with two single roots, the inner surface gravity is nonzero as long as the black hole is nonextremal, $r_+\neq r_-$. These situations are then unstable, as shown in previous work~\cite{Brown:2011tv,Carballo-Rubio:2018pmi,Carballo-Rubio:2021bpr}, with the instability timescale being controlled by the value of $\kappa_-$.

However, the condition $\kappa_-=0$ can be also satisfied away from extremality if $r_-$ is not a single root. This is tantamount to requiring that
\begin{equation}\label{eq:fvan}
   \left. \frac{d F(r)}{dr}\right|_{r=r_-}=0.
\end{equation}
Note that the function $\phi(r)$, which must be finite, can only change the value of $\kappa_-$ when the latter is nonzero. As we are interested in situations in which $\kappa_-=0$ due to Eq.~\eqref{eq:fvan} being satisfied, we can assume that $\phi(r)=0$ for simplicity.

While we have little knowledge of the behaviour of the black hole metric in the immediate vicinity of the inner core, it is reasonable to assume that general relativity holds as a good approximation about the outer horizon $r=r_+$ if the two scales $r_-$ and $r_+$ are well separated. Therefore, we maintain that $r_+$ is a single root of $F(r)$, in analogy with general relativity. In this situation, for the inner surface gravity to vanish, the root $r=r_-$ has to be at least cubic. The lowest possible degree for the polynomials in $F(r)$ to satisfy this requirement is $n=4$. For instance, we would have that {$N_{n=4}(r)$} is given by
\begin{equation}
   {N_{n=4}(r)} =\left(r-r_-\right)^3(r-r_+),
\end{equation}
so that we can write
\begin{equation}
\label{E:F(r)-1}
    F(r)=\frac{\left(r-r_-\right)^3(r-r_+)}{a_4r^4+a_3r^3+a_2r^2+a_1r+a_0}\,.
\end{equation}

Note that, while the coefficients of {$N_{n=4}(r)$} are 
determined {in terms of $r_\pm$ }, this is not true \emph{a priori} for the coefficients of $D_{n=4}(r)$. However, 
the regularity conditions in Eq.~\eqref{eq:metreg}, which are equivalent to
\begin{equation}
    F(r)=1+\mathcal{O}(r^2)\,,
\end{equation}
and the asymptotic condition
\begin{equation}\label{eq:asym}
    F(r)=1-\frac{2M}{r}+\mathcal{O}(r^{-2}),
\end{equation}
can be used to fix
\begin{equation}
\label{E:F(r)-2}
    F(r)=\frac{\left(r-r_-\right)^3(r-r_+)}{
    \left(r-r_-\right)^3(r-r_+) + 2M r^3 + b_2 r^2}\,.
\end{equation}
This specifies the denominator in terms of the physically meaningful quantities $r_-$, $r_+$, $M$, and one remaining free parameter $b_2$, where now $a_2 = b_2 + 3 r_-(r_-+r_+)$.
Thence 
\begin{eqnarray}
    D_4(r) &=& \left(r-r_-\right)^3(r-r_+) + 2M r^3 + b_2 r^2
    \nonumber\\
    &=&
    r^4 + (2M-3r_--r_+) r^3 + a_2 r^2 - r_-^2 (3r_++r_-) r + r_-^3 r_+.
\end{eqnarray}
This can be rewritten as a sum of squares
\begin{eqnarray}
    D_4(r) &=& 
    r^2 \left[r +{(2M-3 r_--r_+)\over 2}\right]^2 +c_2r^2+ 
   r_-^3 r_+\left[1- {1\over2} r\left({3\over r_-} +{1\over r_+}\right)\right]^2,
\end{eqnarray}
where now
\begin{equation}
 a_2 = b_2 + 3 r_-(r_-+r_+) = c_2 + 
 {(2M-3 r_--r_+)^2\over4} + 
 {r_-^3 r_+ (3/r_- +1/r_+)^2\over 4}.
\end{equation}
Certainly, as long as $c_2\geq 0$, (that is, as long as $a_2$, [or $b_2$], are sufficiently large), the denominator will be a positive sum of squares, and so the denominator will have no zeros on the real axis --- and thence the metric function $F(r)$ will have no poles on the real axis. 

Let us now consider some more specific physically plausible choices of parameters.  
Assume that both
$r_-\ll r_+ \sim 2M$ and more specifically that $ r_-\sim \left|r_+-2M\right|$, (the following analysis changes slightly if we do not assume $r_-\sim \left|r_+-2M\right|$).
Under these conditions
\begin{equation}
    a_2 \sim c_2 + 4 r_-^2 +{9\over 4} r_+ r_- \sim c_2 +{9\over 4} \; r_+ r_-.
\end{equation}
The condition for a non-zero denominator ($c_2>0$) is then equivalent to
\begin{equation}
\label{E:a2gtrsim}
 a_2 \gtrsim {9\over 4} \; r_+ r_-.   
\end{equation}
This is the \emph{only} condition we have to impose on $a_2$ in order to avoid the presence of zeros in the denominator.

In summary, we consider the metric in Eq.~\eqref{eq:genmet} with
\begin{equation}
\label{E:F(r)-3}
    F(r)=\frac{\left(r-r_-\right)^3(r-r_+)}{
    \left(r-r_-\right)^3(r-r_+) + 2M r^3 + [a_2-3r_-(r_++r_-)] r^2}\,,\qquad \phi(r)=0\,,
\end{equation}
subject to
\begin{equation}
    r_-\ll r_+ \sim 2M; \qquad   r_-\sim \left|r_+-2M\right|;
    \qquad 
    a_2 \gtrsim {9\over 4} r_+ r_-. 
\end{equation}
%

\section{Analysis of perturbations: double null shell model}\label{sec:dnull}
Let us now study the stability of the geometry just introduced. We begin by following the analysis of \cite{Dray:1985yt,Redmount,Carballo-Rubio:2018pmi} considering a perturbation constituted by two null shells crossing at a radius $ r_0 $ and we study the backreaction on the geometry as $r_0$ approaches the inner horizon along an outgoing null shell. These shells meet at $r_0$ at a given moment of
time. We can also use null coordinates, which is useful as we are interested in analyzing the behavior of the system when $r_0$ is displaced along a null outgoing curve. Hence, we can take a constant value $u = u_0$, being this value arbitrary but
for the condition that it lies inside the outer horizon, and modify the value of $v$, which means that we can describe the trajectory of the crossing point in terms of the curve $r_0(v)_{u=u_0}$.

\begin{figure}
    \centering
    \includegraphics[scale=.6]{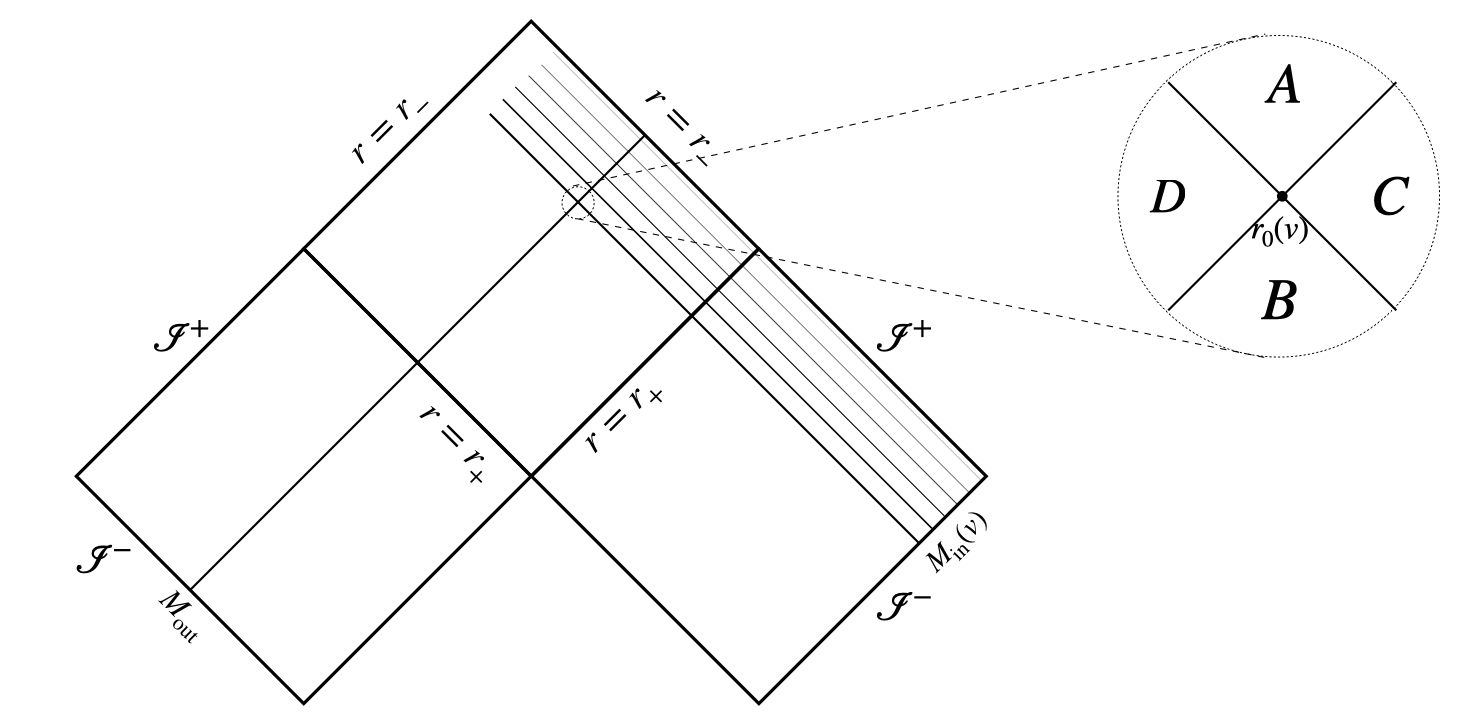}
    \caption{Schematic representation of the relevant sections of the Penrose's diagram of a regular black hole. A pair of outgoing/ingoing null shells cross at a point $r_0(v)$ close to the inner horizon dividing the spacetime into four regions (A, B, C and D). We will consider several ingoing shells and analyze the behavior of the system as the point $r_0(v)$ is displaced along the outgoing null shell.}
    \label{fig:null_shells}
\end{figure}

As shown in Fig.~\ref{fig:null_shells}, we can see that the spacetime is divided into four regions (A, B, C and D).
The DTR relation \cite{Dray:1985yt,Redmount,Carballo-Rubio:2018pmi} allows us to relate the geometry, and hence the Misner--Sharp masses, in the aforementioned four regions, obtaining~\cite{Carballo-Rubio:2018pmi}
\begin{equation}\label{eq:DTR}
    m_A(r_0)=m_B(r_0)+m_{in}(r_0)+m_{out}(r_0)-\frac{2m_{out}(r_0)m_{in}(r_0)}{r_0F_B(r_0)}, 
\end{equation}
where $m_A$ and $m_B$ are the values of the Misner--Sharp mass in the regions A and B respectively, and $m_{out}\equiv m_D-m_B$ and $m_{in}\equiv m_C-m_B$ denote the jump of the Misner-Sharp mass across the two shells. The values of these jumps depend on the energy of the perturbation and on the specific theory under consideration. The quantity $F_B(r)$ in the denominator of Eq.~\eqref{eq:DTR} goes to zero for $r$ close to the inner horizon as 
\begin{equation}\label{eq:FBr}
    F(r)=\frac{1}{3!}F'''(r)\left(r-r_-\right)^3+\mathcal{O}\left[\left(r-r_-\right)^4\right]\,.
\end{equation}
However, before concluding that this leads to an unbounded growth of $m_A$, we need to study the behavior of $m_{in}$ and $m_{out}$. It is useful to impose boundary conditions on the asymptotically flat regions, in particular on $\mathscr{I}^+$ and $\mathscr{I}^-$, and it is useful to use the value of the Bondi mass on $\mathscr{I}^-$, $M_{in}$, to write these conditions explicitly. The Bondi mass will shift due to the existence of an ingoing shell. We can then translate this shift into the shift of the Misner--Sharp mass $m(r)$, which is a function of $r$ and $M_{in}$. At first order, and along the ingoing shell, we would have for instance
\begin{equation}\label{eq:ql_jump_in}
m_{in}(r_0,M,M_{in})=\left.\frac{\partial m}{\partial M}\right|_{r=r_0} M_{in}+\mathcal{O}(M_{in}^2)\,.
\end{equation}
Note the explicit dependence of $m_{in}$ on $r_0$ , while $M_{in}$ has no such dependence. We need to evaluate this dependence explicitly in order to draw conclusions. In physical terms, the meaning of this dependence is the following: a given ingoing shell has a fixed value of $M_{in}$ by construction, but the value of the Misner--Sharp mass changes along the shell, and also changes if the shell is dropped earlier or later (due to the spacetime being time-dependent). Hence, for the setting that we are considering here, in which $r_0$ is displaced along an outgoing null shell, the value of the Misner--Sharp mass at $r_0(v)_{u=u_0}$ will ultimately be a function of $v$.

However, it is useful to evaluate Eq.~\eqref{eq:ql_jump_in} for a fixed value of $v=v_0$ first, and then generalize it to include this time dependence explicitly. For the geometry under consideration, we have
\begin{equation}
    m(r, M)=\frac{1}{2}r\left(1-\frac{\left(r-r_-(M)\right)^3(r-r_+(M))}{
    \left(r-r_-(M)\right)^3(r-r_+(M)) + 2M r^3 + b_2(M) r^2}\right)\,.
\end{equation}
A direct computation of the partial derivative with respect to $M$ shows that the resulting expression can be arranged in two separate pieces:
\begin{equation}\label{eq:dmdM}
    \frac{\partial m}{\partial M}=d_1(r,M)\;\frac{\partial r_-}{\partial M}\; \left(r-r_-(M)\right)^2+d_2(r,M)\; \left(r-r_-(M)\right)^3\,,
\end{equation}
where $d_1(r,M)$ and $d_2(r, M)$ are functions of $r$ and $M$, with the exact functional forms being irrelevant for the current discussion, as we only need to consider their values at the inner horizon, which are given by
\begin{equation}
    d_1\left(r_-,M\right)=-\frac{3}{2}\; \frac{r_+(M)-r_-(M)}{r_-(M)\left(2Mr_-^2(M)+b_2(M)\right)}\,,
\end{equation}
and
\begin{equation}
d_2\left(r_-(M),M\right)=\frac{1}{2}\;\; \frac{\partial}{\partial M}\left(\frac{r_+(M)-r_-(M)}{r_-(M)\; (2Mr_-(M)+b_2(M))}\right)\,.
\end{equation}
The important feature in Eq.~\eqref{eq:dmdM} is that the first term vanishes quadratically when $r_0\rightarrow r_-$ if $\partial r_-/\partial M\neq 0$, while the second term vanishes at least cubically. Hence, the partial derivative $\partial m/\partial M$ vanishes quadratically if the location of the inner horizon depends on the asymptotic mass, which we will assume below as this is the case for the most commonly considered metrics \cite{Carballo-Rubio:2018pmi}.\footnote{{From the discussion of this section it also follows that if $\partial m/\partial M=0$ in a region containing the inner horizon there would be no backreaction of the shells on the geometry. Alternatively, the geometry close to the inner horizon is controlled only by the regularization scale, with no interplay with the ADM mass. Due to the lack of backreaction on the position of the inner horizon, mass inflation cannot manifest in these specific situations.}}

Combining Eqs.~\eqref{eq:dmdM} and~\eqref{eq:ql_jump_in}, we obtain
\begin{equation}\label{eq:ql_jump_in_r}
m_{in}(r_0,M,M_{in})=d_1(r_-,M)\left.\frac{\partial r_-}{\partial M}\right|_{r=r_-} \left(r_0-r_-\right)^2M_{in}+\mathcal{O}[\left(r_0-r_-\right)^3]\,.
\end{equation}
A similar statement applies to $m_{out}$ as a function of $M_{out}$ (the asymptotic mass defined on the other asymptotic region in Fig.~\ref{fig:null_shells}), so that we can write
\begin{equation}\label{eq:ql_jump_in_r}
m_{out}(r_0,M,M_{out})=d_1(r_-,M)\left.\frac{\partial r_-}{\partial M}\right|_{r=r_-} \left(r_0-r_-\right)^2M_{out}+\mathcal{O}[\left(r_0-r_-\right)^3]\,.
\end{equation}

The $r_0$ dependence in these equations translates into a dependence on $v$ when the crossing radius $r_0$ approaches the inner horizon along the outgoing shell, following a trajectory $r_0(v)$. Let us calculate this dependence explicitly. To this end, we note that, along an outgoing null trajectory,
\begin{equation}
    dv=\frac{2\;dr}{F(r)}=\frac{12\;dr}{F'''(r_-)(r-r_-)^3}\left[1+\mathcal{O}\left(r-r_-\right)\right],
\end{equation}
where we have used Eq.~\eqref{eq:FBr}. The leading order of this equation can be easily integrated, obtaining
\begin{equation}\label{eq:r0v}
    v\simeq-\frac{6}{F'''(r_-)(r-r_-)^2}+v_\star\quad\implies\quad    r-r_-\simeq\sqrt{\frac{\left|F'''(r_-)\right|}{v-v_\star}}\,.
\end{equation}
Aside from the dependence on $v$ through $r_0(v)$, in the specific problem we are analyzing $M_{in}$ becomes a function of $v$, as we are considering a stream of ingoing shells (recall Fig.~\ref{fig:null_shells}). Putting all these ingredients together we obtain, at leading order,
\begin{equation}\label{eq:ql_jump_in_r}
m_{in}\left(r_0(v),M,M_{in}(v)\right)\simeq \frac{h(M)}{v}\;M_{in}(v),
\end{equation}
where $h(M)$ is a function of $M$ (thus with no dependence on $v$), and
\begin{equation}\label{eq:ql_jump_in_r}
m_{out}\left(r_0(v),M,M_{out}\right)\simeq \frac{h(M)}{v}\;M_{out}.
\end{equation}
We just need one further ingredient to extract the asymptotic behavior of Eq.~\eqref{eq:DTR}, which is obtained by plugging Eq.~\eqref{eq:r0v} into Eq.~\eqref{eq:FBr}, thus obtaining
\begin{equation}
F(r(v))\simeq \left|F'''(r_-)\right|^{5/2}\;v^{3/2}
\,.
\end{equation}
Substituting these expressions into Eq.~\eqref{eq:DTR}, we obtain 
\begin{equation}
    m_A(r_0(v))=m_B(r_0(v))+h v^{-1}M_{in}(v)+h v^{-1}M_{out}-\frac{2h^2}{r_-\left|F'''(r_-)\right|^{5/2}} v^{-1/2} M_{in}(v)M_{out}.
\end{equation}
Thus we see that the geometry under consideration does not suffer from the  mass inflation instability when perturbed with two null shells. In fact the energy of the perturbations $M_{out}$ and $M_{in}$ is not blueshifted close to the inner horizon, and their backreaction on the geometry goes to zero for large $v$ when $M_{in}(v)$ decays in time, or even in the case in which it remains constant.

\section{Analysis of perturbations: Modified Ori model}\label{sec:ori}
\begin{figure}
    \centering
    \includegraphics[scale=.65]{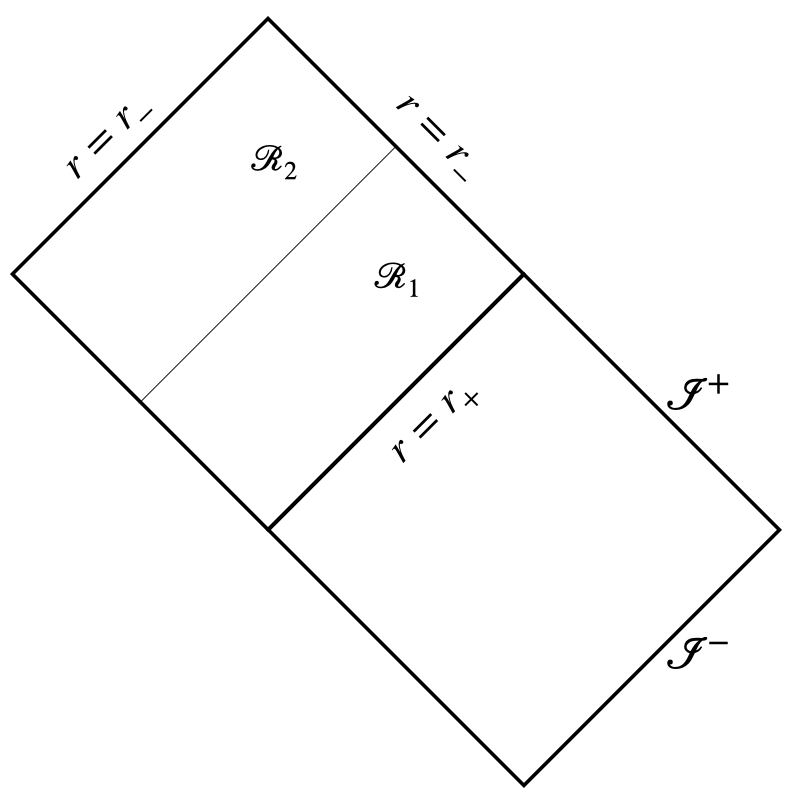}
    \caption{Schematic representation of the relevant sections of the Carter--Penrose diagram of a regular black hole. An outgoing null shell divides the spacetime in two regions
    \label{Fig:Ori}}
\end{figure}

Let us now consider a different model for perturbations in which the ingoing null geodesic is replaced by a continuous flux of matter. This setup was originally considered by Ori in \cite{Ori:1991zz} to study the instability of the Cauchy horizon of a Reissner--Nordstr\"om black hole. The outgoing shell, which is a shared ingredient with the model studied in the previous section, divides the spacetime into two regions. With reference to Fig.~\ref{Fig:Ori}  we denote the region exterior to this shell as $\mathcal{R}_{1}$, while the interior region will be $\mathcal{R}_{2}$.
The metrics in each of these regions can be written without loss of generality as a generalization of Eq.~\eqref{eq:genmet} in which the functions $F(r)$ and $\phi(v)$ are  now time-dependent:
\begin{equation}\label{eq:orimet}
      \text{d}s^2_{1/2}=-e^{-2\phi_{1/2}(v,r)}F_{1/2}(v,r)\text{d}v^2+2e^{-\phi(r)}\text{d}v\text{d}r+r^2\text{d}\Omega^2.
\end{equation}
As we have seen in the previous section, and is discussed in numerous works~\cite{Poisson1989,Ori:1991zz,Hamilton2008,Brown:2011tv,Carballo-Rubio:2018pmi,Bertipagani:2020awe,Bonanno:2020fgp,Carballo-Rubio:2021bpr}, the instability is controlled by the inner surface gravity, the main features of which are contained in the function $F(v,r)$. Hence, for simplicity we can consider that $\phi(v,r)=\phi(v)$, which can be then absorbed in a redefinition of $v$. In practice, we will thus be working with the metrics
\begin{equation}\label{eq:orimetu}
      \text{d}s^2_{1/2}=-F_{1/2}(v,r)\text{d}v^2+2\text{d}v\text{d}r+r^2\text{d}\Omega^2.
\end{equation}

We will further assume that both $F_{1}$ and $F_{2}$ take the functional form in Eq.~\eqref{E:F(r)-3}, with $M_{1/2}$ being promoted to functions of $v$. The functional form of $M_{1}$ determines the amount of energy that is being absorbed by the black hole. There are different choices for this function and, to follow standard conventions, we will assume that the decay of the perturbation is described by the power law
\begin{equation}\label{eq:price}
    M_{1}=M_0-\beta \left(\frac{v}{v_0}\right)^{-\gamma}.
\end{equation}
This relation is typically used at late times, in which regime it is known as Price's law~\cite{Price1971,Price1972}. While focusing only on the late-time behavior is justified in situations in which there is mass inflation or, in other words, to argue for unstable behavior, arguing for stability would require one to show that quantities remain bounded at all points during evolution. Hence, we will assume that Eq.~\eqref{eq:price} is valid at all times as a working assumption.

Eq.~\eqref{eq:price} fixes the form of the metric in the region external to the shell, which means that we just need to determine the behavior of internal metric. In particular, we are interested in understanding the behavior of the gravitational energy enclosed by the outgoing shell, which is given by $m_{2}(v,r_0)$, which is related to $F_{2}(v,r_0)$ by Eq.~\eqref{eq:msdef}. This function can be obtained using the junction conditions on the outgoing shell which, as described in~\cite{Carballo-Rubio:2021bpr}, results into the equation
\begin{equation}\label{eq:modOri}
       \left.\frac{1}{F_{1}}\frac{\partial m_{1}}{\partial v}\right|_{r=R(v)}= \left. \frac{1}{F_{2}}\frac{\partial m_{2}}{\partial v}\right|_{r=R(v)}\,,
\end{equation}
in which $v:=v_{1}$ is the $v$ coordinate in the exterior region.\\
This equation can be solved numerically assuming that the geometry is described by Eq.~\eqref{E:F(r)-3} both outside and inside the shell. To perform the numerical integration, we need to make some choices on the free parameters. As a working example, we consider
\begin{equation}
    r_+=2M;\qquad\qquad r_-=\ell \left( 1+\alpha \frac{\ell}{M}+\mathcal{O}\left(\frac{\ell^2}{M^2}\right)\right).
\end{equation}
where $\ell$ is the regularization parameter and $\alpha$ is an order one constant. The reason for this choice of the expression for $r_-$ is that it describes the location of the inner horizon for the geometries that are usually studied in the literature \cite{Carballo-Rubio:2018pmi}.

\begin{figure}
    \centering
    \includegraphics[scale=.45]{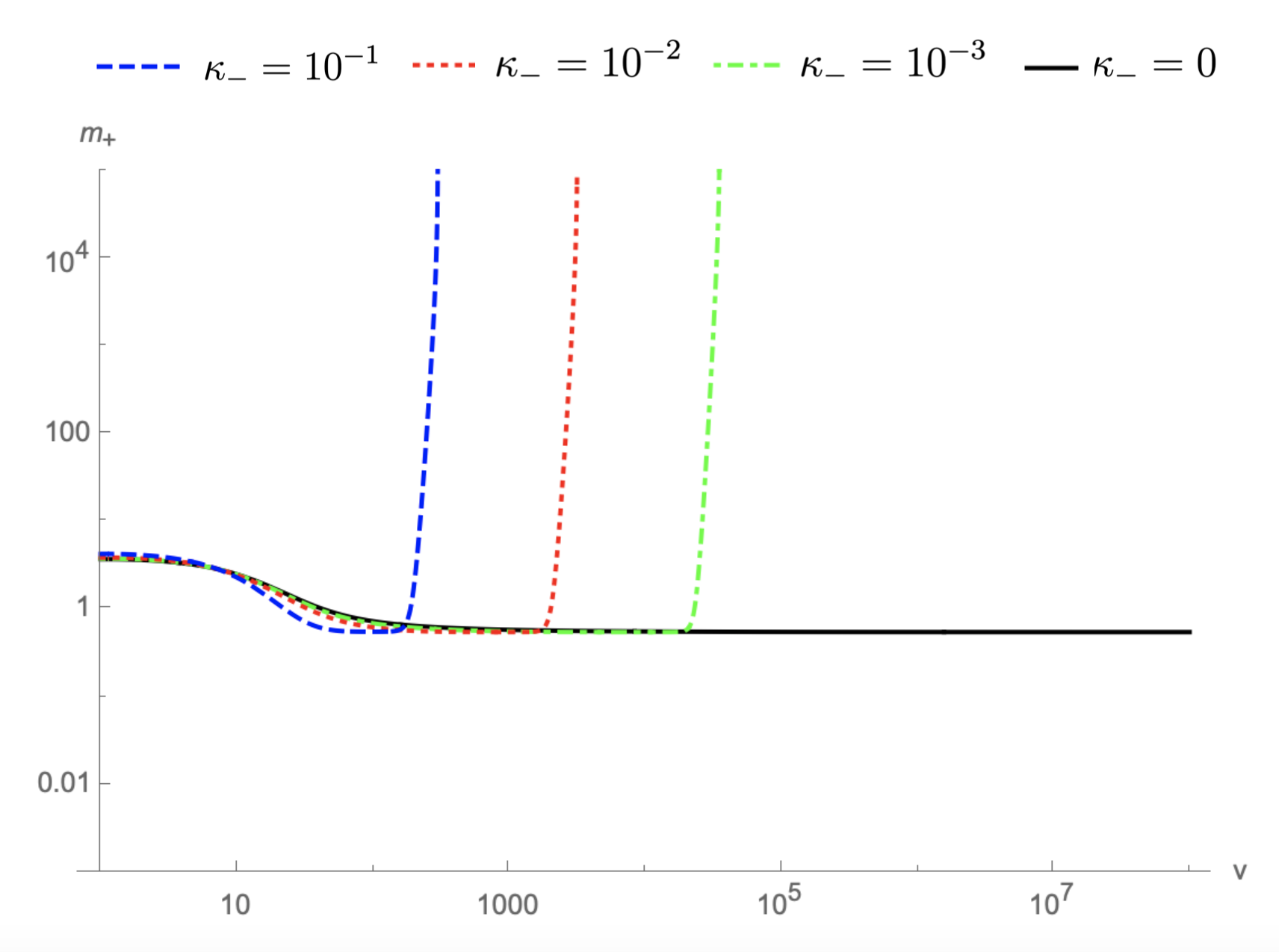}
    \caption{Numerical evolution of the Misner--Sharp $m_+$ for different values of $\kappa_-$. For the numerical integration we have considered $M_0=100$, $\ell=1$, $\alpha=\beta=v_0=1$, $\gamma=12$, $a_2=10M_0\ell$.
    \label{fig:my_label}}
\end{figure}

Finally, in order to check that the absence of the instability is due the the vanishing of the surface gravity at the inner horizon, we will study the slightly modified  geometry
\begin{equation}
        F(r)=\frac{(r-r_-)(r-kr_-)^2 (r-r_+)}{r^4+(2 M-3 r_--r_+)r^3+a_2r^2- r_-^2
   (r_-+3 r_+)r+r_-^3 r_+} \,.
\end{equation}
For $k=1$ this geometry reduces to the one in \eqref{E:F(r)-3}, whereas for $k\neq1$ the surface gravity at the inner horizon is non-zero, and is given by
\begin{equation}
    \kappa_-=\frac{1}{2}\frac{dF}{dr}=-\frac{r_+-r_-}{2\left(2Mr_-+a_2-3r_-\left(r_-+r_+\right) \right)}\left( 1-k \right)^2\,.
\end{equation}
Fig.~\ref{fig:my_label} shows the result of the numerical integration of Eq.~\eqref{eq:modOri}. We can see that the instability timescale is longer for smaller values of the surface gravity, and for $\kappa_-=0$ the numerical integration does not show any sign of instability. 

\section{Conclusions} \label{sec:conc}

We have discussed the features of a new kind of regular black hole that combines features of non-extremal and extremal black holes: inner-extremal regular black holes have two horizons, at positions $r=r_+$ and $r=r_-$, an arbitrary outer surface gravity $\kappa_+$, and a vanishing inner surface gravity $\kappa_-=0$. 

The main motivation behind our proposal is the fact that, in previously analyzed regular black in which $\kappa_-\neq0$, the timescale for mass inflation (and the corresponding exponential instablity) is controlled by $\kappa_-$. Hence, that it is possible to find regular black holes in which the inner surface gravity vanishes, suggests that these situations can avoid the mass inflation issue.

However, answering this question was not as simple as using the expressions for $\kappa_-\neq0$ and taking the limit $\kappa_-\rightarrow0$, as these expressions are not valid in this limit. Reaching a definitive answer required evaluating the next order in the equations evolving the evolution of perturbations on top of these geometries. We have performed this calculation for two different models that are routinely used in the study of mass inflation: a model with two null shells, and a model with a null shell and a continuous stream of radiation (Ori model). In both cases, we have seen that there is indeed no exponential mass inflation, and that the backreaction of perturbations vanishes asymptotically as long as perturbations decay in time (or, at most, remain constant).

These non-extremal regular black holes with $\kappa_-=0$ are therefore natural candidates to consider as alternative to singular black holes, and also to regular black holes with $\kappa_-\neq 0$. Our proposal represents an improvement with respect to these two kinds of geometries, and provides a proof of principle that singularity regularization does not need to result in exponential instabilities. A fundamental question that remains to be answered is whether these geometries could represent the stable endpoint of the dynamical evolution driven by the backreaction of perturbations in unstable ($\kappa_-\neq0$) regular black holes. We hope to come back to this question in the future.

\vspace{-10pt}
\acknowledgments 
\enlargethispage{60pt}
\vspace{-10pt}
RCR acknowledges financial support through a research grant (29405) from VILLUM fonden.
FDF acknowledges financial support by Japan Society for the Promotion of Science Grants-in-Aid for international research fellow No. 21P21318. 
SL acknowledges funding from the Italian Ministry of Education and  Scientific Research (MIUR)  under the grant  PRIN MIUR 2017-MB8AEZ. 
CP acknowledges the financial support provided under the European Union's H2020 ERC, Starting Grant agreement no.~DarkGRA--757480 and support under the MIUR PRIN and FARE programmes (GW- NEXT, CUP: B84I20000100001).
MV was supported by the Marsden Fund, via a grant administered by the Royal Society of New Zealand. 
\clearpage
\bibliography{refs}
\end{document}